\documentclass[sigchi,manuscript]{acmart}

\usepackage{natbib}
\usepackage{booktabs} % For formal tables
\usepackage[utf8x]{inputenc} 
\usepackage{balance}
\usepackage{titlesec}
\usepackage{multirow}
\usepackage{graphicx}

\def\BibTeX{{\rm B\kern-.05em{\sc i\kern-.025em b}\kern-.08emT\kern-.1667em\lower.7ex\hbox{E}\kern-.125emX}}
\copyrightyear{2020}
\acmYear{2020}
\setcopyright{acmlicensed}\acmConference[MobileHCI '20]{22nd International Conference on Human-Computer Interaction with Mobile Devices and Services}{October 5--8, 2020}{Oldenburg, Germany}
\acmBooktitle{22nd International Conference on Human-Computer Interaction with Mobile Devices and Services (MobileHCI '20), October 5--8, 2020, Oldenburg, Germany}
\acmPrice{15.00}
\acmDOI{10.1145/3379503.3403563}
\acmISBN{978-1-4503-7516-0/20/10}

\begin{document}

%
% The "title" command has an optional parameter, allowing the author to define a "short title" to be used in page headers.
\title[Comparing IPA use for native and non-native language speakers]{See what I'm saying? Comparing Intelligent Personal Assistant use for Native and Non-Native Language Speakers}

%
% The "author" command and its associated commands are used to define the authors and their affiliations.
% Of note is the shared affiliation of the first two authors, and the "authornote" and "authornotemark" commands
% used to denote shared contribution to the research.
\author{Yunhan Wu}
\affiliation{University College Dublin}
\email{yunhan.wu@ucdconnect.ie}

\author{Daniel Rough}
\affiliation{University College Dublin}
\email{daniel.rough@ucd.ie}

\author{Anna Bleakley}
\affiliation{University College Dublin}
\email{anna.bleakley@ucdconnect.ie}

\author{Justin Edwards}
\affiliation{University College Dublin}
\email{justin.edwards@ucdconnect.ie}

\author{Orla Cooney}
\affiliation{University College Dublin}
\email{orla.cooney@ucdconnect.ie}

\author{Philip R. Doyle}
\affiliation{University College Dublin}
\email{philip.doyle1@ucdconnect.ie}

\author{Leigh Clark}
\affiliation{Swansea University}
\email{l.m.h.clark@swansea.ac.uk}

\author{Benjamin R. Cowan}
\affiliation{University College Dublin}
\email{benjamin.cowan@ucd.ie}

%
% By default, the full list of authors will be used in the page headers. Often, this list is too long, and will overlap
% other information printed in the page headers. This command allows the author to define a more concise list
% of authors' names for this purpose.
% \renewcommand{\shortauthors}{}

%
% The abstract is a short summary of the work to be presented in the article.
\begin{abstract}
Limited linguistic coverage for Intelligent Personal Assistants (IPAs) means that many interact in a non-native language. Yet we know little about how IPAs currently support or hinder these users. Through native (L1) and non-native (L2) English speakers interacting with Google Assistant on a smartphone and smart speaker, we aim to understand this more deeply. Interviews revealed that L2 speakers prioritised utterance planning around perceived linguistic limitations, as opposed to L1 speakers prioritising succinctness because of system limitations. L2 speakers see IPAs as insensitive to linguistic needs resulting in failed interaction. L2 speakers clearly preferred using smartphones, as visual feedback supported diagnoses of communication breakdowns whilst allowing time to process query results. Conversely, L1 speakers preferred smart speakers, with audio feedback being seen as sufficient. We discuss the need to tailor the IPA experience for L2 users, emphasising visual feedback whilst reducing the burden of language production. 

\end{abstract}


%
% Keywords. The author(s) should pick words that accurately describe the work being
% presented. Separate the keywords with commas.

%
% A "teaser" image appears between the author and affiliation information and the body 
% of the document, and typically spans the page. 

%
% This command processes the author and affiliation and title information and builds
% the first part of the formatted document.
 
\maketitle
\fancyhead{}

\section{Introduction}
The proliferation of voice based Intelligent Personal Assistants (IPAs) like Google Assistant, on smart speakers and smartphones, has made speech a common interaction modality \cite{clark2019state}. Although many IPAs can now be used in languages other than English, coverage and supported functionality is by no means comprehensive, varying by device and assistant (e.g., \cite{kinsella_2019}). This creates a barrier for those whose first language is not fully supported, forcing them to interact in a non-native language or face being excluded from IPA use. Prominent work in HCI looks at IPA user experience \cite{cowan_what_2017, luger_like_2016, porcheron_voice_2018} almost exclusively from the perspective of first language (L1) English speakers, leaving the experience of users who engage with IPAs in a non-native language (such as L2 speakers) unclear.  

The work presented contributes important insight into how L2 speakers experience IPAs. Our work aims to 1) map significant dimensions of L2 user experience whilst also 2) identifying how aspects of the two most popular devices for IPA use (smartphones and smart speakers \cite{kinsella_2018})  support or hinder L2 users. We also compare this to L1 speaker experiences so as to emphasise the contrasting needs of these speaker groups. To achieve this we carried out a study where, following interactions with Google Assistant on both a smartphone and on a smart speaker,  L1 and L2 English speakers took part in a semi-structured interview devised to gain insight into their experiences. Our results highlight a number of clear differences between L1 and L2 speakers' perceptions and experiences of IPA use. We found L1 and L2 speakers varied in their interaction approaches, whereby L2 speakers focused heavily on their pronunciation as opposed to L1 emphasising the need for simple, succinct and well planned utterances. Both emphasised the need for speech adaptation, informed by perceived system limitations, yet L2 speakers' adaptation were also driven by their own perceived linguistic limitations. Whereas L1 speakers felt the IPA waited too long to speak after they gave a command, L2 speakers felt the assistant was not sensitive enough to the extra time they needed to produce their command and process the system's utterances. This resulted in the system regularly interrupting L2 speakers. L2 speakers consistently expressed the desire for IPA design to support lexical retrieval or reduce the need for language production, yet this was not a concern for L1 speakers. We also discovered a clear difference in device preference across the speaker groups. L2 speakers significantly preferred using IPAs on smartphones, because they provided visual feedback that supported their interaction. In contrast, L1 speakers preferred smart speakers, with audio feedback being seen as sufficient to support interaction. Our findings build on recent interest in L2 speakers \cite {pyae_18, pyae_19}, contributing a deeper insight into L2 IPA experiences across these device types. Our findings emphasise the need to consider L2 user needs so as to be more inclusive of this group. Our work suggests that tailoring IPA interaction by being sensitive to the time needed by L2 users, concentrating on visual feedback and reducing the need for language production in interaction are key design priorities to support L2 users.

\section{Related Work}
\subsection{Interacting with Intelligent Personal Assistants}
Recent research in HCI has predominantly focused on understanding IPA user experience from the L1 perspective. It highlights that a major benefit of speech as a modality lies in its facilitation of multitasking, especially in hands-busy/eyes-busy situations such as driving or looking after children \cite{luger_like_2016, cox_tlk} (although the benefits of multitasking with speech are heavily dependent on primary task demand \cite{edwards_cui_2019_interruptions}). IPAs are commonly used for information search, controlling music applications, setting alarms and timers, and to control IoT (Internet of Things) devices (e.g. smart lights) \cite{tawfig_kaye_TOCHI}, through limited question-answer type dialogues \cite{gilmartin_social_2017, porcheron_voice_2018}. Especially through smart speakers, these assistants can be used by multiple people at once, becoming a social focal point \cite{porcheron_voice_2018}. Although clearly useful, previous work has also identified a number of issues.  These include users not trusting IPAs to execute  more complex or socially sensitive tasks (e.g. sending a message or calling a contact) \cite{luger_like_2016}, perceived problems in accurately recognising accented speech \cite{cowan_what_2017}, and the humanlike design of IPAs poorly signalling actual IPA capabilities \cite{cowan_what_2017,luger_like_2016,doyle_humanness_2019, weiss2020}. Privacy and data collection practices are also a concern for users \cite{cowan_what_2017, clark_what_2019}, driven particularly by the always-listening nature of devices, along with concerns about how speech data is used and stored and who has access to it \cite{tawfig_kaye_TOCHI}.

\subsection{Non-native speakers' use of IPAs}
Although languages can be changed and added on a number of popular IPAs, coverage and functionality varies across devices and assistants (e.g.\cite{kinsella_2019}). This forces many users to have to speak to IPAs in a language other than their first language. The amount of work on IPA use among L2 speakers is limited, with research being preliminary and questionnaire based in nature \cite{pyae_18}, or focused on L2 language learning technology experience \cite {dizon2017using, alemi2015impact}. Recent quantitative research suggests that L2 speakers find smart speakers harder to use \cite {pyae_18, pyae_19} and more difficult to interact with effectively than L1 speakers \cite {pyae_18, pyae_19}, with language proficiency being related to more positive experience ratings \cite {pyae_19}.  Although they enjoy the interaction, L2 speakers feel they have to expend considerable effort when planning their utterances \cite {pyae_19}. Rephrasing commands also leads L2 speakers to become frustrated \cite {pyae_19}. IPA use has also been explored as a tool to help L2 users improve language skills \cite{ipa_learning_l2} as they afford L2 speakers an opportunity to practise listening to speech output and produce speech input in a stress free context \cite{moussalli2016commercial}. This work has noted that, although they may not be perceived as such \cite{cowan_what_2017}, IPAs are adept at recognising accented speech accurately in these contexts, whilst providing a useful tool for L2 language learners \cite{ipa_learning_l2}. Work has also observed how non-native speech output, particularly the matching of accents with non-native users, can significantly impact user perceptions and behaviour. Non native speakers were more dissatisfied with speech outputs when they used accents that varied from their own \cite{dahlback2001spoken} whilst non-native speakers also failed to respond to information provided by spoken navigation systems in accents that were dissimilar to their own \cite{jonsson2011can}.

\subsection{Non-native speaker interaction experiences}
Non-native speaker’s interaction experiences, such as their experiences of web page readability and internet search, have also been studied, revealing more general HCI difficulties L2 speakers may face. Web page readability studies have shown that vocabulary retrieval and parsing of complex grammatical structures are two major difficulties for L2 speakers when reading English pages \cite{yu2010enhancing, yu2011enhancing}. Likewise, a study of online search found L2 users to have particular difficulties in query formation due to both vocabulary limitations and grammatical phrasing. This work also found that repair strategies such as rephrasing are much more difficult for L2 users and add to technology-related stress \cite{chu2015online}. While these difficulties are found in non-native use of written language in technology use, we expect that L2 users may experience similar challenges in speech based interactions with technology. 

\section{Research aims}
Most IPA user experience research has focused on identifying issues in L1 speaker interactions. Recent work on L2 speakers has used more quantitative approaches to explore their experiences. Our works builds on this through using qualitative techniques to more deeply investigate the experiences of L2 speakers. To this end, our paper presents research that aims to identify important issues in L2 IPA user experience, emphasising these issues through a contrast with L1 speaker experiences. In particular, we aim to explore how characteristics of the two most popular IPA device platforms, namely: smartphones and smart speakers \cite{kinsella_2018} influence these experiences. Our work is the first to directly compare L1 and L2 users’ experiences of an off-the-shelf IPA across these device platforms. Through a deepened understanding of challenges faced by L2 users, we hope to inform IPA design toward more inclusive interactions.

\section{Method}
\subsection{Participants}
Thirty three participants (F=14, M=18, Prefer not to say=1; Mean age=28.1 yrs; SD=9.8 yrs) were recruited from a European university via email, posters and flyers displayed across campus, and through snowball sampling. One participant was removed from the sample due to technical issues in their experiment session. Of the remaining thirty two participants, 16 (F=8, M=7, Prefer not to say=1) were native English speakers, with English as their first language, and 16 (F=6, M=10) were native Mandarin speakers who were non-native English speakers. On a 7 point Likert Scale (1= Not at all proficient; 7 = Extremely proficient) our sample of 16 Mandarin speakers rated their English proficiency as moderate (Mean=4.21, SD=0.7). Across the sample 78.1\%  (N=25) had used IPAs before, with 9.4\% (N=3) reporting frequent or very frequent use. Among participants that reported previous experience with IPAs, 13 were native Mandarin speakers and 12 were native English speakers, with Siri (56\%) being most commonly used, followed by Amazon Alexa (36\%) and Google Assistant (12\%). 

\subsection{Device conditions}
In the experiment, participants interacted with Google Assistant, through both a Moto G6 smartphone (\emph{Smartphone} condition) and a Google Home Mini smart speaker (\emph{Smart speaker} condition) using a within participants design. The order in which these were experienced was counterbalanced across L1 and L2 speaker groups. Using Google Assistant across both conditions ensured that the devices were the only source of difference. Google Assistant was selected because it is commonly used across both device types \cite{olson_2019_2019}.   

\subsection{Task}
Participants were asked to conduct a total of 12 tasks with Google Assistant (six per device - all tasks are included in supplementary material). Based on research identifying the most common tasks people conduct with IPAs \cite{tawfig_kaye_TOCHI, dubiel_survey_2018} experimental tasks included 1) playing music, 2) setting an alarm, 3) converting values, 4) asking for the time in a particular location, 5) controlling device volume and 6) requesting weather information. Two versions of each task were generated creating two sets of six tasks. Each set of 6-tasks were used in only one of the device conditions. All tasks were delivered to participants as pictograms (see Figure 1). This was so as to eliminate the potential influence of written task instructions on what both L1 and L2 participants might say to the IPAs, to more closely simulate natural query generation and to reduce potential difficulties translating task text for L2 speakers. The task sets were counterbalanced across device and speaker conditions and task order was randomised within sets for each participant.  

\begin{figure*}[t!]
    \centering
    \includegraphics[keepaspectratio, width=0.6\textwidth]{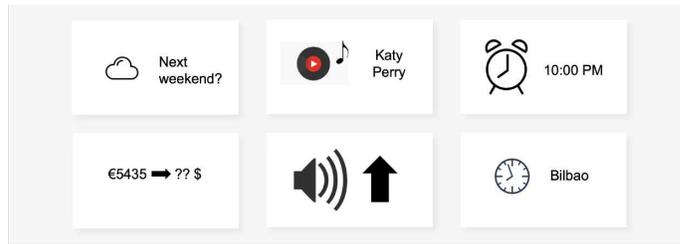}
    \caption{Example set of task pictograms}
    \label{fig:tasks}
\end{figure*}

\subsection{Post interaction interview}
After interacting with both devices, participants took part in a semi-structured interview. These interviews lasted approximately 20 minutes and focused on 3 key topics: 1) general views towards IPAs; 2) experiences with the IPAs in the experiment; and 3) reflections on how they spoke to each system. Participants were also asked to identify which of the devices they preferred and explain their preference. So as to ensure that there were no linguistic barriers when expressing their opinions, L2 speaker interviews were conducted in Mandarin. All data was audio recorded and transcribed, with the Mandarin interviews being translated back into English by a Mandarin speaking member of the research team. The interview data was then analysed using inductive thematic analysis \cite{braun2006using}. Initial coding was conducted independently by two of the research team with experience in qualitative data analysis using thematic approaches. After initial codes and themes were generated, these were discussed by both coders and further refined based on these discussions. The results of this analysis are presented in the Results section below.

\subsection{Procedure}
The research received ethical approval through the University's low risk project ethics procedures [HS-E-19-127]. English and Mandarin speakers were recruited from staff and students at a European University via email, posters and through snowball sampling. Upon arriving at the lab, participants were fully briefed about the nature of the study before being asked to provide written consent. Next, they were asked to complete a demographic questionnaire, giving information about their age, sex, profession, nationality, native language and English proficiency. Both groups were asked to give details regarding whether they had used IPAs before, and if so how frequently. They were also asked to select which they had used most frequently from a list of IPAs (Siri, Alexa, Google Assistant, Microsoft Cortana, Samsung Bixby or other). 

Participants were then introduced to the task pictograms, before interacting with Google Assistant. In order to ensure participants did not encounter significant difficulties interpreting the pictograms, they were first presented with a trial set of images (on paper) and asked to write down what they would say to the agent to complete each task. These were similar in topic and layout to the pictograms used during their smartphone and smart speaker Google Assistant interactions, but varied in the information being requested (e.g. varying the city for which the time is requested). Participants' interpretations were checked by the experimenter prior to interacting with the devices. 

After completing this, participants were presented with a set of six pictograms on a laptop, one at a time, and asked to complete the task represented in the image using Google Assistant on one of the devices. Participants self-reported the completion of tasks when they either thought they had accomplished the goal of the task or felt stuck and unable to complete the task. Tasks were deemed complete by participants rather than experimenters in order to avoid influencing interaction strategies. Once participants felt they had completed a task, they were asked to confirm by clicking a checkbox on the screen. By clicking a button, participants then revealed the next task. This process was then repeated until all six tasks were completed for that device. The participant was then asked to complete a further six tasks with Google Assistant using the next device. The tasks were delivered using the same process. An experimenter was present, only to ensure the tasks were being engaged with and to help with any technical issues. 

After finishing the tasks with both smartphone and smart speaker devices, participants then completed a short post-interaction interview. This was recorded using a Blue Yeti microphone and audio capture software (Audacity v. 2.3.0). After completing the interview, participants were fully debriefed as to the aims of the study, uses of the data they had provided, and were given contact information for any further questions. They were then thanked for participation and given a €10 voucher as an honorarium. The study lasted approximately 40 minutes. 

\section{Results}

\subsection{General speaker differences}
Irrespective of device type, clear differences emerged between L1 and L2 speakers' experiences when using IPAs. These revolved around how each group approached the interaction, issues in turn taking, and the desire for design to reduce the effort involved in language production.

\subsubsection{Interaction approaches:}

Echoing previous literature on language production in speech interface \cite{kennedy_dialogue_1988} and IPA \cite{cowan_what_2017, porcheron_voice_2018, luger_like_2016} use, L1 speakers prioritised vocal clarity, using correct English, brevity and planning when approaching interaction with both IPAs: 

\begin{center}
\textit{``I suppose it’s breaking it down and thinking about what the question actually is. I suppose what it is you want to know, how you should ask that if you were using proper English.''}~[P17-L1] 
\end{center}

\begin{center}
\textit{``[A]rticulate clearly and you know, think about how was, what was the simplest way possible to ask the question you know.''}~[P4-L1]    
\end{center}

L2 speakers also heavily emphasised adaptations aimed at increasing the likelihood of being understood, but this was due to their own perceived language or speech limitations. For instance, rather than adapting vocabulary to try and improve system performance (as is commonly mentioned in previous work \cite{luger_like_2016, cowan_what_2017}) L2 speakers instead altered their vocabulary based on whether or not they knew a particular word:

\begin{center}
\textit{``I tend to change the words I used. For example, when I asked the devices to change the volume, I didn't know the word ‘volume’ so I changed it to ‘voice’ or ‘sound’.''}~[P9-L2]
\end{center}

L2 speakers tried to adopt strategies to overcome this, especially when the IPAs repeatedly did not recognise their utterances, although these were unsuccessful: 

\begin{center}
\textit {``The recognition of people’s names and place names has a low success rate. And when I don’t know a word, I would spell it. But the devices cannot understand that.''}~[P9-L2]
\end{center}

L2 speakers were also highly sensitive to their pronunciation or need to retrieve the correct words in interaction, which took up considerable time when interacting with the IPAs:

\begin{center}
\textit{''... sometimes it cannot understand some pronunciation by non-native English speakers, and it cannot help you pronounce the words you don’t know.''}~[P9-L2]
\end{center}

\begin{center}
    \textit{``As a non-native English speaker, I have a hard time on proper nouns ... Maybe the devices cannot recognize due to this issue, and it may waste a lot of time.''}~[P4-L2]
\end{center}

\subsubsection{Waking and turn taking:}
There were also clear differences between L1 and L2 speakers when it came to \textit{waking} the device and managing turn taking with the IPAs. L2 speakers regularly felt like they struggled to wake the Assistants in both device conditions: 

\begin{center}
\textit {``I feel that I wake up a few times and it ignored me ... like you have finished the commands but it maybe didn’t get you.''}~[P14-L2]
\end{center}

During interactions, L2 speakers suggested they sometimes needed extra time to formulate an utterance and that this was not taken into consideration by the system. Subsequently, the system would either reset or barge in before they had finished their request:

\begin{center}
\textit{``I need some time to think ... if I think for a long time, I have to wake up the machine again. The command I said before is wasted.''}~[P4-L2]
\end{center}

\begin{center}
\textit{``[W]hen I didn’t finish my sentences and the machine `thinks' that I’m finished. It’s like, the IPAs can only analyze the sentences you said.''}~[P14-L2]
\end{center}

In contrast, L1 speakers perceived the delay between speaking to the device and it responding as too long, making the interaction seem slow:

\begin{center}
\textit{``[A]gain it would be the delayed interaction. What I think is going on there is it is trying to figure out whether or not you finished a sentence?''}~[P9-L1]
\end{center}

\begin{center}
\textit{``It’s too slow to react, because it can’t keep up with me. I’d much rather type...''}~[P1-L1]
\end{center}

This led L1 speakers to question whether the devices were working correctly, increasing frustration:  

\begin{center}
\textit{``[T]here were times when I wasn’t sure if you know it was just not working or something? Because I really clearly said what I wanted it to do and it didn’t, you know, react.''}~[P4-L1]
\end{center}

\begin{center}
\textit{``It’s more frustrating I suppose, the lack of proper response to it, there was no, ‘don’t understand the questions’ ... the feedback didn’t seem adequate.''}~[P17-L1]
\end{center}

\subsubsection{Reducing the burden of production:}
L2 speakers also tended to emphasise ways that IPAs could be improved, with a focus on reducing the level of language production needed as well as ways to support word recall and gaps in lexical knowledge. Frustration with having to reproduce or reformulate queries from scratch was common, with L2 speakers suggesting that it would be helpful if IPAs were aware of previous attempts to make a query:

\begin{center}
\textit{``It’s, maybe I can meet some words problems. We are not native speakers after all. So I hope that it can, you, know… recognize this part more intelligently. For example, I use some simple words to describe my commands and the devices can understand the meaning. The devices can try to figure out my requirements. That will be better.''}~[P7-L2]  
\end{center}

\begin{center}
\textit{``After I finished the question, for example, when I asked the time of a city, and got an answer for a wrong city, then there is no need to ask the whole question again. I should only need to emphasize the city name.''}~[P1-L2]
\end{center} 

Similarly, to support L2 issues with word selection, participants suggested that contextual options could be provided in cases where the intent of the query was recognised but users were struggling with the specific noun required:

\begin{center}
\textit{``When I don't know that word I can say `please transfer Celsius degree to another temperature unit'...for example, they can list all temperatures for you.''}~[P8-L2]
\end{center}

\begin{center}
\textit{``For example, if you say some unclear words and the devices can show you the possible choices or match the most possible option or something.''}~[P16-L2]
\end{center}

This desire to support the interaction with suggestions was not identified by L1 users, who instead suggested better recognition of wider forms of language such as colloquialisms: 

\begin{center}
\textit{``[T]hey're interesting I suppose in how they try to make you speak in a different way that’s not natural to you. They make your colloquialism sound strange and they make you pronounce things in that curious kind of way.''}~[P17-L1]
\end{center}

\begin{center}
\textit{``Improved? I’m not too sure, I suppose it involves listening to more conversations and getting a colloquial idea than just proper English.''}~[P5-L1]
\end{center}

\subsection{Device-specific differences}
Our analysis discovered a marked difference in the way that both speaker types experienced IPA use through the smart speaker and smartphone respectively. There was a clear difference in preference between the speaker groups, driven by the benefits of visual feedback and output for supporting the interaction. 

\subsubsection{L1 and L2 device preference:}
A Chi-squared test revealed a statistically significant difference in device preference between L1 and L2 speakers [ $\chi^2(1,N=30)=4.74, p < 0.05$]. Among the L1 speakers, 75\% (N=12) preferred using Google Assistant on the smart speaker, whilst 63\% of L2 speakers (N=10) preferred using the Assistant on the smartphone. Two participants reported having no preference. Preference frequencies are shown in Table \ref{table1}. 

\begin{table}
\caption{Frequency of participant device preferences}
\label{table1}
\begin{tabular}{ |c|c|c|c| } 
\hline
Group & Smart Speaker & Smartphone & No Preference \\
\hline
L1 & 12 & 4 & 0 \\ 
\hline
L2 & 4 & 10 & 2 \\
\hline
\end{tabular}

\end{table}

\subsubsection{Visual confirmation:}

Both L1 and L2 speakers commented on the role of visual feedback in IPA interactions using a smartphone. However, L2 speakers placed much greater emphasis on the benefits of this feature in allowing them to build confidence in the efficacy of the system’s speech recognition capabilities, and in supporting them to identify exact reasons for miscommunication. Having visual feedback on the smartphone when using the Assistant also reduced the burden of having to interpret, translate and retain information for L2 speakers, which was an issue when responses and feedback were solely delivered using speech (i.e. when using the smart speaker).
As shown in the \textit {General Speaker Differences} section, it was common for L2 speakers to doubt whether they had pronounced words correctly. The on-screen feedback with the smartphone allowed them to identify exactly what the system did and did not recognise:

\begin{center}
\textit{``When I interacted with the second one [smartphone], I feel good. Because sometimes I know that my pronunciation is not accurate. But it can recognize the words accurately.''}~[P7-L2]
\end{center}  

\begin{center}
\textit{``I feel I didn’t pronounce accurately for some location names. However, the interface can recognize my pronunciation correctly.''}~[P6-L2]
\end{center} 

In addition to building user confidence in the system's recognition capabilities, this transparency also allowed non-native speakers to localise specific lexical items that were the cause of miscommunication:

\begin{center}
\textit{``However, for the smartphone, for example, I don’t know how to ask the question for temperature. Then I asked ‘How to describe the temperature in two systems’ and the system replied to me. Although I still cannot pronounce those two words, I can receive some information about that. The smartphone can show the text on the screen.''}~[P9-L2]
\end{center}

In line with this observation, L2 speakers noted having the opposite experience with the smart-speaker:

\begin{center}
\textit{``For the first interface [smart speaker], I even have no idea about where I got it wrong. It’s like...maybe you have misarticulation, however, you never know which words you have the pronunciation issues with'.'}~[P3-L2]
\end{center}

\subsubsection{Visual output:}
In addition to transcriptions of the speech being recognised by the IPA, L2 speakers also benefited from supplementary information, displayed on-screen in the smartphone condition, in response to queries. Here, the pairing of speech and visual output from the system was found to alleviate difficulties in interpreting audio feedback whilst simultaneously trying to retain the information given by the system:

\begin{center}
\textit{``[A]s a non-native English speaker, reading is much easier than listening. That device [smartphone] has a screen for people to read the text. I can gather more information using that.''}~[P4-L2]
\end{center}

\begin{center}
\textit{``If I didn’t pay attention on hearing the answers I can check the details on the screen.''}~[P10-L2]
\end{center}

Whilst the visual output provided in the smartphone condition was viewed positively by a number of L2 speakers, it was often negatively perceived by L1 speakers. These users stated that it reduced the usefulness of the experience when visual output was provided as an alternative to speech-based responses:

\begin{center}
 \textit{``[I]t brought up a list of web resources and I thought I can do that myself, you know. I expected it to give me a response rather than leave me to look at sites where I can get the information myself''}~[P14-L1]
\end{center} 

\begin{center}
\textit{''I suppose, it’s easy enough to check things on your phone anyways so like, I don’t feel I need to say it. Like I don’t know.''}~[P2-L1]
\end{center}

\section{Discussion}
\phantomsection
IPAs are useful, especially in facilitating interaction in hands busy/eyes busy contexts \cite{cowan_what_2017, luger_like_2016}. Although language coverage has recently been broadened to accommodate non-English speaking users, functionality and device coverage across these languages is highly variable (e.g. \cite{kinsella_2019}), excluding these users from the full benefits of IPA use. Building on recent preliminary questionnaire-based work \cite{pyae_18, pyae_19}, our research, the first to compare IPA experiences across devices for L1 and L2 users, identifies in detail the key aspects of L2 user experience, how this compares to L1 users, and whether this varies across smartphone and smart speaker based IPA interactions. Through thematic analysis, we identified a number key themes that highlight differences between L1 and L2 users. We found clear differences in the approaches taken to language production when generally interacting with the IPAs. L1 speakers perceived themselves to focus on structuring their commands as succinctly and as simply as possible. Although L2 speakers also tended to plan and adapt their natural speech, they focused more on how this was driven by perceived language limitations, in particular pronunciation, lexical knowledge and retrieval. Poor IPA waking and turn taking was also a problem experienced by L2 speakers. They felt that Google Assistant was not sensitive to the time they needed to produce their utterances, meaning they regularly experienced system barge-in. In contrast, L1 speakers felt that the time between production and recognition was too long, disrupting the interaction.  When considering ways the interaction with IPAs might be improved, L2 speakers consistently emphasised approaches that would reduce the need for language production or support lexical retrieval. This was not a concern aired by L1 speakers. Our work adds richer detail to our current understanding of the difficulties faced by L2 users in IPA interactions \cite{pyae_18, pyae_19} along with a deeper understanding of differing potential causes of difficulty and frustration faced by L1 and L2 users respectively.

Our research also discovered a clear difference in device preference between L1 and L2 speakers, with L2 speakers preferring smartphone based IPA interactions. This was in contrast to L1 speakers, who preferred using the smart speaker. In particular, L2 speakers found the visual feedback provided through the smartphone critical in supporting their interaction, helping them diagnose reasons for interaction breaking down, whilst also giving them confidence in the system's ability to understand their utterances. L1 speakers, on the other hand, felt that audio-only feedback was sufficient. This also extends to system output, whereby L2 speakers benefited from the display of query results through text or through onscreen information, with the L1 speakers highlighting this as unnecessary. Thus, our work highlights how accessing IPAs through screen-based rather than speech-only devices, can be useful in supporting L2 speakers; reducing both the level of effort required for successful interaction and the level of frustration they experience. We explore the reasons and design implications of these findings below.

\subsection{Designing IPAs to be sensitive to L1 and L2 interaction differences}
Our findings highlight some important differences between how L1 and L2 speakers interact with IPAs. L1 speakers emphasised the importance of succinct and short utterances. This supports previous observations of user behaviour, whereby users tend to simplify or adapt their language choices when interacting with speech interfaces \cite{kennedy_dialogue_1988, amalberti_user_1993,luger_like_2016} to increase the likelihood of interaction success \cite{oviatt_linguistic_1998, branigan_role_2011, cowan2019s}. This is thought to be driven by users seeing speech interfaces as \textit{at risk listeners} \cite{oviatt_linguistic_1998} or poor interlocutors \cite{branigan_role_2011}, whereby adaptation to perceived system limitations is required to ensure interaction success. Although L2 speakers may also perceive speech systems in the same way, L2 speakers seemed to more heavily place the burden of potential interaction failure on themselves, seeing their pronunciation and lack of linguistic knowledge as significant barriers. This should be considered when designing IPA interactions. L2 users may need to be given more time to produce utterances, and more opportunities to clarify perceived misinterpretations of speech output, requiring multiple turns to repair and negotiate miscommunications \cite{hoekje_processes_1984}. These are not currently afforded by the one size fits all approach of current IPAs. Future design of IPAs should look to tailor the experience if the system identifies a user as a non-native speaker.  Without these changes, L2 speakers may be at risk of abandoning IPA use more readily. Recent work focusing on language learning contexts, suggests that abandonment, along with direct repetition and rephrasing of queries, are common L2 strategies when faced with miscommunication with IPAs \cite{ipa_learning_l2}. Future research should more deeply explore ways to tailor the IPA experience based on this, and how that may influence long term IPA engagement.

\subsection{Screens are integral for L2 IPA users}
Screen-based feedback was clearly important in supporting L2 users' IPA experience. For instance, speech recognition transcriptions displayed on screen were found to play a role in developing L2 speakers' confidence in the system's recognition capabilities, whilst also helping them to diagnose specific reasons for communication breakdown. Using the screen to support speech output from the system with supplementary information, such as links to websites or maps, was also seen as a non-trivial benefit among L2 users. Previous work illuminates potential reasons for this. L2 speakers find non-native synthesis less intelligible than L1 speakers \cite{reynolds1996synthetic, mayasari_l2_synthesis}, particularly in noisy environments \cite{reynolds1996synthetic}. The interpretation of non-native speech also significantly increases cognitive load for L2 interlocutors during dialogue interactions \cite{segalowitz2005automaticity,dornyei1998problem} as linguistic dimensions (e.g. sound system and common linguistic structures) may vary significantly from their native language \cite{watson2013effect}. Although reading non-native language text is also likely to increase cognitive load,  the permanence of supplementary visual feedback \cite{ipa_learning_l2}, may give L2 users extra time to process and comprehend the information and refer back to it at a later time, improving their experience. This is not possible with speech only smart speakers. Research on the cognitive implications of screens in voice user interface (VUI) design is needed to support this further, yet from our study it is clear that incorporating well designed screen based feedback is a design imperative for improving L2 speaker interaction with IPAs. Future studies should build on this, by concentrating on identifying specific ways that this screen based feedback can be improved to further support L2 speakers.

\subsection{The need to relieve the burden of production}
L2 speakers also regularly mentioned difficulties in producing what they felt were the correct words, or pronunciations of words, to interact effectively with the IPA. Suggestions for IPA improvements echoed these difficulties. For instance, it was felt that systems should allow the user to correct a single lexical item - the root cause of a miscommunication - whilst preserving the context of a query, or that they should list a small range of options associated with specific common tasks. This was so as to save L2 users the effort of formulating multiple queries for a task. Again, this is likely to be due to the additional difficulties and increase in cognitive load associated with retrieving appropriate lexical items across multiple languages. This retrieval difficulty is because, compared to monolinguals, bilinguals experience less frequent word activation through processing and production \cite{gollan2004tot}, making the word needed at a specific moment hard to access compared to monolingual users \cite{segalowitz2005automaticity}. This makes retrieval for production particularly cognitively taxing \cite{dornyei1998problem}, with our L2 participants suggesting that the IPA needs to  be sensitive to the time this takes. Literature within psycholinguistics and HCI also suggests other ways in which IPAs could be designed to reduce this production burden. For instance, system speech output could be used to increase activation of relevant, in-domain lexicon and syntax, by using priming \cite{cowan2015does, cowan2015voice, branigan_role_2011}; essentially priming  keywords, nouns and structures the system can effectively manage. This could be facilitated through speech only and multimodal interactions and may be particularly useful in error management, assisting L2 speaking users in reformulating queries. Indeed, such techniques may reduce language production load for all users, and thus lead to significant user experience benefits across L1 and L2 users. Future work needs to focus more specifically on how these techniques may influence user experience. 
\subsection{Limitations}
This work examines how L2 speakers of English interact with IPAs, highlighting how this compares to L1 English speakers. To ensure that we could conduct interviews in L2 speakers' first language, all of the L2 participants in the present work were native speakers of Mandarin. These participants were students of [European University] living in a country where English is a primary language and thus are more frequently exposed to English than other L2 English speakers might be. Although this may limit generalisability, our findings may be conservative, as people with less frequent exposure to English would likely have even greater difficulty in interacting effectively in IPAs. Further, the level of dissimilarity between Mandarin and English may also impact the generalisability of our findings. English and Mandarin vary significantly on a number of dimensions (e.g. relative importance of structure and tone in defining meaning, frequency of abstract and concrete nouns \cite{devescovi2004competition}). L2 speakers with a more similar first language to English (e.g. native Germanic language speakers) may vary in their experiences compared to the L2 users in this study. Future work should look to explore this in more detail. In addition, our work uses Google Assistant for both interactions. Future work in this topic should look to include a wider and more diverse selection of L2 speakers, as well as explore L2 experiences with other assistants and contexts.

To compare participant experiences with smartphone and smart speaker based systems, all participants interacted with both in each experiment. Through these interactions they completed two sets of six tasks, with similarities in the way they were graphically presented. Participants' experiences with the IPA through the first device may of course influence their interaction with the second device, as they may use similar commands or alter commands based on their initial experience. They may also be less hesitant in engaging with the IPA and completing the tasks second time around. To reduce the potential for practice effects we ensured that the task sets were randomised and were counterbalanced across the experiment conditions. Likewise, device conditions were also counterbalanced within groups.

Rather than using text or audio to deliver the tasks, we constructed a set of 12 pictograms, with two versions of each task. This was specifically used to reduce the likelihood of participants using the exact wording used in task instructions had the instructions been written or spoken. Using pictograms, although adding another cognitive demand on users through the need to interpret the image, was designed to encourage users to generate queries in a way that is more in-keeping with `real-world' interactions. The use of text in this experiment might also have unduly increased the cognitive load of L2 speakers as they would have to translate the task text and then generate the query. The pictogram method used meant they could interpret the images without the need for task translation. So as to ensure that these were interpreted accurately, we also made participants report how they would word a query to the IPA for that particular image, with the experimenter ensuring that these were accurately interpreted and any issues in interpretation could be clarified before commencing the experiment. Due to the nature of the participants in this type of work, it is important for future work to consider the nature of task delivery and the potential interference that using linguistic means to communicate tasks may have on the findings of future research.

\section{Conclusion}
 IPAs, although accessible to native English speakers, are not universally accessible in all languages. Language coverage varies by device and/or assistant. This means it may not be feasible for all users to interact in their native tongue. Our study has outlined key themes related to how L1 and L2 speakers vary in their user experience, and how aspects of IPAs may benefit or impede L2 users. We find that using IPAs through smartphones, which afford visual feedback to support the user, are significantly preferred by L2 users. We also identified important differences in the strategies L2 and L1 users had when planning their utterances to ensure communication success, with L2 users looking to identify ways to reduce their levels of language production in interaction. Importantly, by comparing L1 and L2 users, the work highlights specific areas that may be leveraged to support L2 speakers in future IPA use. It also demonstrates the importance of expanding the types of users being researched to ensure that IPAs are designed to be more inclusive and accessible to a wider audience.

% The acknowledgments section is defined using the "acks" environment (and NOT an unnumbered section). This ensures
% the proper identification of the section in the article metadata, and the consistent spelling of the heading.
\begin{acks}
This work was conducted with the financial support of the UCD China Scholarship Council (CSC) Scheme grant No. 201908300016, Science Foundation Ireland ADAPT Centre under Grant No. 13/RC/2106 and the Science Foundation Ireland Centre for Research Training in Digitally-Enhanced Reality (D-REAL) under Grant No. 18/CRT/6224.
\end{acks}

%
% The next two lines define the bibliography style to be used, and the bibliography file.
\balance
\bibliographystyle{ACM-Reference-Format}
\bibliography{L2}

\end{document}